\documentclass[aps,preprint,tightenlines,superscriptaddress,showpacs,byrevtex]{revtex4-1}

\usepackage{graphicx}
\usepackage{epsfig}
\usepackage{dcolumn}
\usepackage{bm}
\usepackage{color}
\usepackage{amsmath}

\usepackage{ulem}
\usepackage{siunitx}
\usepackage{lineno}
\DeclareSIUnit\barn{b}

\newcommand{\fb}{\si{\femto\barn}}
\newcommand{\infb}{\fb^{-1}}
\newcommand{\pb}{\si{\pico\barn}}

\newcommand{\gpsp}{\gamma\psp}
\newcommand{\gisr}{\gamma_{\rm ISR}}

\newcommand{\gev}{\rm GeV}
\newcommand{\gevc}{{\rm GeV}/c}
\newcommand{\gevcs}{{\rm GeV}/c^2}
\newcommand{\mev}{\rm MeV}

\newcommand{\mevcs}{{\rm MeV}/c^2}
\newcommand{\kev}{\rm keV}

\newcommand{\ev}{\rm eV}

\newcommand{\MMS}{M_{\rm rec}^2}

\newcommand{\yfos}{\Upsilon(4S)}
\newcommand{\yns}{\Upsilon(nS)}

\newcommand{\eff}{\varepsilon}
\newcommand{\BR}{{\cal B}}
\newcommand{\jpc}{J^{PC}}

\newcommand{\psp}{\psi(2S)}
\newcommand{\chicpt}{\chi_{c2}(2P)}

\newcommand{\jpsi}{J/\psi}

\newcommand{\EE}{e^+e^-}
\newcommand{\MM}{\mu^+\mu^-}
\newcommand{\LL}{\ell^+\ell^-}
\newcommand{\GG}{\gamma\gamma}
\newcommand{\pp}{\pi^+\pi^-}

\newcommand{\ddb}{D\overline{D}}

\newcommand{\ppjpsi}{\pi^+\pi^- J/\psi}

\newcommand{\cm}{\si{\centi\metre}}

\newcommand{\reduline}{\bgroup\markoverwith
{\textcolor{red}{\rule[0.5ex]{2pt}{0.4pt}}}\ULon}

\newcommand{\beq}{\begin{equation}}
\newcommand{\eeq}{\end{equation}}
\newcommand{\bitm}{\begin{itemize}}
\newcommand{\eitm}{\end{itemize}}

\def\Journal#1#2#3#4{{#1} {\bf #2}, #3 (#4)}

\def\NIMA{Nucl. Instrum. Methods A}
\def\NPB{Nucl. Phys. B}
\def\PLB{Phys. Lett. B}
\def\PRL{Phys. Rev. Lett.}
\def\PRD{Phys. Rev. D}
\def\PRP{Phys. Rep.}
\def\RMP{Rev. Mod, Phys.}
\def\PTEP{Prog. Theor. Exp. Phys.}

\def\EPJC{Eur. Phys. J. C}
\def\JHEP{JHEP}

\begin{document}

\preprint{} \preprint{ \vbox{ \hbox{   }   
        \hbox{Belle Preprint 2021-08}
        \hbox{KEK Preprint 2021-4}
        }}  

\title{\quad\\[2.0cm]
 The Study of $\GG\to\gpsp$ at Belle}

%% %simple case: 2 authors, same institution
%% \author{A. Uthor}
%% \author{and A. Nother Author}
%% \affiliation{Institution,\\Address, Country}

% more complex case: 4 authors, 3 institutions, 2 footnotes
%\author{X.~L.~Wang,\note{Corresponding author.}}
%\author{B.~S.~Gao,}
%\author{W.~J.~Zhu}

% The "\note" macro will give a warning: "Ignoring empty anchor..."
% you can safely ignore it.

%\affiliation[a]{Fudan University,\\220 Handan Rd., Yangpu District, 200433, Shanghai}

% e-mail addresses: one for each author, in the same order as the authors
%\emailAdd{xiaolong@fudan.edu.cn}
%\emailAdd{gaobs@ihep.ac.cn}
%\emailAdd{zhuwj@ihep.ac.cn}

%%% Paper:    g g -> g psi(2S)
%%% Journal:  JHEP
%%% Contacts: X.L. Wang (xiaolong@fudan.edu.cn)
%%%           B.S. Gao (gaobs@ihep.ac.cn)
%%%           W.J. Zhu (zhuwj@ihep.ac.cn)
%%% Non-responding authors or those who said NO are commented out.
%%% ====================================================================
%%% Click the RELOAD button on your web browser to see the updated file.
%%% ====================================================================
%%% Use \input{pubxxx} to insert this material into your latex file.
%%% Special instructions for JHEP:
%%%  1) add ",\hbox{$\dagger$}" inside the affiliation [...] and
%%%         "\note[$\dagger$]{Corresponding author.} after the corresponding author's name.
%%%  2) add  \emailAdd{...}" for the corresponding author at the end of this file.
\noaffiliation
\affiliation{Department of Physics, University of the Basque Country UPV/EHU, 48080 Bilbao, Spain}
\affiliation{University of Bonn, 53115 Bonn, Germany}
\affiliation{Brookhaven National Laboratory, Upton, New York 11973, USA}
\affiliation{Budker Institute of Nuclear Physics SB RAS, Novosibirsk 630090, Russian Federation}
\affiliation{Faculty of Mathematics and Physics, Charles University, 121 16 Prague, The Czech Republic}
\affiliation{Chonnam National University, Gwangju 61186, South Korea}
\affiliation{University of Cincinnati, Cincinnati, OH 45221, USA}
\affiliation{Deutsches Elektronen--Synchrotron, 22607 Hamburg, Germany}
\affiliation{Duke University, Durham, NC 27708, USA}
\affiliation{Department of Physics, Fu Jen Catholic University, Taipei 24205, Taiwan}
\affiliation{Key Laboratory of Nuclear Physics and Ion-beam Application (MOE) and Institute of Modern 
Physics, Fudan University, Shanghai 200443, PR China}
\affiliation{Justus-Liebig-Universit\"at Gie\ss{}en, 35392 Gie\ss{}en, Germany}
\affiliation{Gifu University, Gifu 501-1193, Japan}
\affiliation{II. Physikalisches Institut, Georg-August-Universit\"at G\"ottingen, 37073 G\"ottingen, Germany}
\affiliation{SOKENDAI (The Graduate University for Advanced Studies), Hayama 240-0193, Japan}
\affiliation{Gyeongsang National University, Jinju 52828, South Korea}
\affiliation{Department of Physics and Institute of Natural Sciences, Hanyang University, Seoul 04763, South Korea}
\affiliation{University of Hawaii, Honolulu, HI 96822, USA}
\affiliation{High Energy Accelerator Research Organization (KEK), Tsukuba 305-0801, Japan}
\affiliation{J-PARC Branch, KEK Theory Center, High Energy Accelerator Research Organization (KEK), Tsukuba 305-0801, 
Japan}
\affiliation{Higher School of Economics (HSE), Moscow 101000, Russian Federation}
\affiliation{Forschungszentrum J\"{u}lich, 52425 J\"{u}lich, Germany}
\affiliation{IKERBASQUE, Basque Foundation for Science, 48013 Bilbao, Spain}
\affiliation{Indian Institute of Science Education and Research Mohali, SAS Nagar, 140306, India}
\affiliation{Indian Institute of Technology Bhubaneswar, Satya Nagar 751007, India}
\affiliation{Indian Institute of Technology Guwahati, Assam 781039, India}
\affiliation{Indian Institute of Technology Hyderabad, Telangana 502285, India}
\affiliation{Indian Institute of Technology Madras, Chennai 600036, India}
\affiliation{Indiana University, Bloomington, IN 47408, USA}
\affiliation{Institute of High Energy Physics, Chinese Academy of Sciences, Beijing 100049, PR China}
\affiliation{Institute for High Energy Physics, Protvino 142281, Russian Federation}
\affiliation{Institute of High Energy Physics, Vienna 1050, Austria}
\affiliation{INFN - Sezione di Napoli, 80126 Napoli, Italy}
\affiliation{INFN - Sezione di Torino, 10125 Torino, Italy}
\affiliation{J. Stefan Institute, 1000 Ljubljana, Slovenia}
\affiliation{Institut f\"ur Experimentelle Teilchenphysik, Karlsruher Institut f\"ur Technologie, 76131 Karlsruhe, 
Germany}
\affiliation{Kavli Institute for the Physics and Mathematics of the Universe (WPI), University of Tokyo, Kashiwa 
277-8583, Japan}
\affiliation{Kennesaw State University, Kennesaw GA 30144, USA}
\affiliation{Department of Physics, Faculty of Science, King Abdulaziz University, Jeddah 21589, Saudi Arabia}
\affiliation{Kitasato University, Sagamihara 252-0373, Japan}
\affiliation{Korea Institute of Science and Technology Information, Daejeon 34141, South Korea}
\affiliation{Korea University, Seoul 02841, South Korea}
\affiliation{Kyoto Sangyo University, Kyoto 603-8555, Japan}
\affiliation{Kyungpook National University, Daegu 41566, South Korea}
\affiliation{Universit\'{e} Paris-Saclay, CNRS/IN2P3, IJCLab, 91405 Orsay, France}
\affiliation{P.N. Lebedev Physical Institute of the Russian Academy of Sciences, Moscow 119991, Russian Federation}
\affiliation{Faculty of Mathematics and Physics, University of Ljubljana, 1000 Ljubljana, Slovenia}
\affiliation{Ludwig Maximilians University, 80539 Munich, Germany}
\affiliation{Luther College, Decorah, IA 52101, USA}
\affiliation{Malaviya National Institute of Technology Jaipur, Jaipur 302017, India}
\affiliation{Faculty of Chemistry and Chemical Engineering, University of Maribor, 2000 Maribor, Slovenia}
\affiliation{Max-Planck-Institut f\"ur Physik, 80805 M\"unchen, Germany}
\affiliation{School of Physics, University of Melbourne, Victoria 3010, Australia}
\affiliation{University of Mississippi, University, MS 38677, USA}
\affiliation{University of Miyazaki, Miyazaki 889-2192, Japan}
\affiliation{Moscow Physical Engineering Institute, Moscow 115409, Russian Federation}
\affiliation{Graduate School of Science, Nagoya University, Nagoya 464-8602, Japan}
\affiliation{Kobayashi-Maskawa Institute, Nagoya University, Nagoya 464-8602, Japan}
\affiliation{Universit\`{a} di Napoli Federico II, 80126 Napoli, Italy}
\affiliation{Nara Women's University, Nara 630-8506, Japan}
\affiliation{National Central University, Chung-li 32054, Taiwan}
\affiliation{National United University, Miao Li 36003, Taiwan}
\affiliation{Department of Physics, National Taiwan University, Taipei 10617, Taiwan}
\affiliation{H. Niewodniczanski Institute of Nuclear Physics, Krakow 31-342, Poland}
\affiliation{Nippon Dental University, Niigata 951-8580, Japan}
\affiliation{Niigata University, Niigata 950-2181, Japan}
\affiliation{University of Nova Gorica, 5000 Nova Gorica, Slovenia}
\affiliation{Novosibirsk State University, Novosibirsk 630090, Russian Federation}
\affiliation{Osaka City University, Osaka 558-8585, Japan}
\affiliation{Pacific Northwest National Laboratory, Richland, WA 99352, USA}
\affiliation{Panjab University, Chandigarh 160014, India}
\affiliation{Peking University, Beijing 100871, PR China}
\affiliation{University of Pittsburgh, Pittsburgh, PA 15260, USA}
\affiliation{Punjab Agricultural University, Ludhiana 141004, India}
\affiliation{Research Center for Nuclear Physics, Osaka University, Osaka 567-0047, Japan}
\affiliation{Meson Science Laboratory, Cluster for Pioneering Research, RIKEN, Saitama 351-0198, Japan}
\affiliation{Department of Modern Physics and State Key Laboratory of Particle Detection and Electronics, 
University of Science and Technology of China, Hefei 230026, PR China}
\affiliation{Seoul National University, Seoul 08826, South Korea}
\affiliation{Showa Pharmaceutical University, Tokyo 194-8543, Japan}
\affiliation{Soochow University, Suzhou 215006, China}
\affiliation{Soongsil University, Seoul 06978, South Korea}
\affiliation{Sungkyunkwan University, Suwon 16419, South Korea}
\affiliation{School of Physics, University of Sydney, New South Wales 2006, Australia}
\affiliation{Department of Physics, Faculty of Science, University of Tabuk, Tabuk 71451, Saudi Arabia}
\affiliation{Tata Institute of Fundamental Research, Mumbai 400005, India}
\affiliation{Department of Physics, Technische Universit\"at M\"unchen, 85748 Garching, Germany}
\affiliation{School of Physics and Astronomy, Tel Aviv University, Tel Aviv 69978, Israel}
\affiliation{Toho University, Funabashi 274-8510, Japan}
\affiliation{Department of Physics, Tohoku University, Sendai 980-8578, Japan}
\affiliation{Earthquake Research Institute, University of Tokyo, Tokyo 113-0032, Japan}
\affiliation{Department of Physics, University of Tokyo, Tokyo 113-0033, Japan}
\affiliation{Tokyo Institute of Technology, Tokyo 152-8550, Japan}
\affiliation{Virginia Polytechnic Institute and State University, Blacksburg, VA 24061, USA}
\affiliation{Wayne State University, Detroit, MI 48202, USA}
\affiliation{Yamagata University, Yamagata 990-8560, Japan}
\affiliation{Yonsei University, Seoul 03722, South Korea}

\author{X.~L.~Wang}\affiliation{Key Laboratory of Nuclear Physics and Ion-beam Application (MOE) and Institute of 
Modern Physics, Fudan University, Shanghai 200443, PR China} % Fudan
\author{B.~S.~Gao}\affiliation{Key Laboratory of Nuclear Physics and Ion-beam Application (MOE) and Institute of Modern 
Physics, Fudan University, Shanghai 200443, PR China} % Fudan
\author{W.~J.~Zhu}\affiliation{Key Laboratory of Nuclear Physics and Ion-beam Application (MOE) and Institute of Modern 
Physics, Fudan University, Shanghai 200443, PR China} % Fudan
\author{I.~Adachi}\affiliation{High Energy Accelerator Research Organization (KEK), Tsukuba 305-0801, 
Japan}\affiliation{SOKENDAI (The Graduate University for Advanced Studies), Hayama 240-0193, Japan} % KEK
\author{H.~Aihara}\affiliation{Department of Physics, University of Tokyo, Tokyo 113-0033, Japan} % Tokyo
\author{S.~Al~Said}\affiliation{Department of Physics, Faculty of Science, University of Tabuk, Tabuk 71451, Saudi 
Arabia}\affiliation{Department of Physics, Faculty of Science, King Abdulaziz University, Jeddah 21589, Saudi Arabia} 
% Tabuk
\author{D.~M.~Asner}\affiliation{Brookhaven National Laboratory, Upton, New York 11973, USA}   % BNL
\author{H.~Atmacan}\affiliation{University of Cincinnati, Cincinnati, OH 45221, USA} % Cincinnati
\author{V.~Aulchenko}\affiliation{Budker Institute of Nuclear Physics SB RAS, Novosibirsk 630090, Russian 
Federation}\affiliation{Novosibirsk State University, Novosibirsk 630090, Russian Federation} % BINP
\author{T.~Aushev}\affiliation{Higher School of Economics (HSE), Moscow 101000, Russian Federation} % HSE
% ...
% Lebedev
\collaboration{The Belle Collaboration}

\date{\today}

\begin{abstract}

Using $980~\infb$ of data at and around the $\Upsilon(nS)(n=1,2,3,4,5)$ resonances collected with the Belle detector at 
the KEKB asymmetric-energy $\EE$ collider, the two-photon process $\GG\to \gamma\psp$ is studied from the threshold to 
$4.2~\gev$ for the first time. Two structures are seen in the invariant mass distribution of $\gpsp$: one at $M_{R_1} = 
3922.4\pm 6.5 \pm 2.0~\mevcs$ with a width of $\Gamma_{R_1} = 22\pm 17\pm 4~\mev$, and another at $M_{R_2} = 4014.3\pm 
4.0 \pm 1.5~\mevcs$ with a width of $\Gamma_{R_2} = 4\pm 11 \pm 6~\mev$; the signals are parametrized with the 
incoherent sum of two Breit-Wigner functions. The first structure is consistent with the $X(3915)$ or the 
$\chi_{c2}(3930)$, and the local statistical significance is determined to be $3.1\sigma$ with the systematic 
uncertainties included. The second matches none of the known charmonium or charmoniumlike states, and its global 
significance is determined to be $2.8\sigma$ including the look-elsewhere effect. The production rates are 
$\Gamma_{\GG}\BR(R_1\to\gpsp) = 9.8\pm 3.6\pm 1.2~\ev$ assuming $(\jpc, |\lambda|) =(0^{++}, 0)$ or $2.0\pm 0.7\pm 
0.2~\ev$ with $(2^{++}, 2)$ for the first structure and $\Gamma_{\GG}\BR(R_2\to\gpsp) = 6.2\pm 2.2\pm 0.8~\ev$ with 
$(0^{++}, 0)$ or $1.2\pm 0.4\pm 0.2~\ev$ with $(2^{++}, 2)$ for the second one. Here, the first errors are statistical 
and the second systematic, and $\lambda$ is the helicity. 

\end{abstract}


%\keywords{$\EE$ Experiments, Quarkonium, Exotic States, Two-photon Collision}

\maketitle
%\flushbottom

\section{Introduction}

More than two dozen new resonances that are dubbed as $X$, $Y$ and/or $Z$ states have been found above the $\ddb$ 
threshold since Belle observed the $X(3872)$ (now labeled the $\chi_{c1}(3872)$~\cite{PDG}) in $B\to 
K\ppjpsi$~\cite{x3872}, and this number is much larger than the expectation from predictions of conventional 
quark-antiquark models. Among these, candidates for both conventional and exotic charmonium-like states are discussed 
widely~\cite{review}. Many puzzles arise from these $XYZ$ states, and one of them concerns the candidates for $P$-wave 
triplet states near $3.9~\gevcs$, including the $X(3872)$, $Z(3930) \to\ddb$ and $X(3915)\to\omega\jpsi$ observed in 
two-photon collisions~\cite{z3930_belle, z3930_babar, z3930_lhcb, x3915_belle, x3915_babar}, and $X^*(3860)\to\ddb$ 
observed in a full amplitude analysis of the process $\EE\to\jpsi \ddb$~\cite{x3860_belle}. 

One of the most interesting $XYZ$ states is the $X(3872)$, which lies very near the $D\bar{D}^*+c.c.$ mass threshold and 
is conjectured to have a large $D\bar{D}^*+c.c.$ molecular component~\cite{Swanson}. Its large production rates in $pp$ 
and $p\bar{p}$ collision experiments~\cite{d0-x-ppbar, lhcb-x-pp, cms-x-pp, atlas-x-pp} and the determination of its 
quantum number by LHCb~\cite{lhcb-x-jpc} suggest that there is a conventional charmonium $\chi_{c1}(2P)$ core in its 
wave function. This is supported by another study of $X(3872)\to\gpsp$ by LHCb~\cite{lhcb-x-gpsp}. A study of the 
lineshape of this state by LHCb reveals a pole structure that is compatible with a quasibound state of 
$D^0\bar{D}^{*0}$ but allowing a quasivirtual state at the level of $2\sigma$~\cite{x3872_lineshape}. Partners of the 
$X(3872)$ are suggested, and one of them is a $D^*\bar{D}^*$ loosely bound state with quantum numbers $\jpc = 
2^{++}$~\cite{x2-guofk, x2-width-guofk}. Belle found evidence for $X(3872)$ production in two-photon 
collisions~\cite{belle-x-tp}, thus motivating the search for the possible $2^{++}$ partner of the $X(3872)$ in such 
collisions. Such a study can provide essential information to understand the nature of the $X(3872)$.

Concurrently, there have been many studies related to the $\chi_{cJ}(2P)$ triplet states. The $Z(3930)$ was discovered 
by Belle in the process $\GG\to\ddb$, and the angular distribution was used to identify it as the $\chi_{c2}(2P)$ 
state~\cite{z3930_belle}. The existence of $Z(3930)$ and its angular distribution were confirmed by 
BaBar~\cite{z3930_babar}. The $X(3915)$ was discovered by Belle~\cite{x3915_belle} and a spin-parity analysis of this 
state by BaBar favored the $\jpc = 0^{++}$ quantum numbers~\cite{x3915_babar}. The $X(3915)$ is a candidate of the 
$\chi_{c0}(2P)$ state~\cite{liu-1, liu-2, liu-3, liu-4}. In a recent amplitude analysis of the $B^+\to K^+D^+D^-$ decay 
by LHCb~\cite{ddb-lhcb}, there are both $0^{++}$ and $2^{++}$ states at $m(D^+D^-)\approx 3930~\mevcs$. Their 
parameters are determined to be $M=3923.8\pm 1.5\pm 0.4~\mevcs$ and $\Gamma = 17.4\pm 5.1\pm 0.8~\mev$ for 
$\chi_{c0}(3930)$ and $M = 3926.8\pm 2.4\pm 0.8~\mevcs$ and $\Gamma = 34.2\pm 6.6\pm 1.1~\mev$ for $\chi_{c2}(3930)$. 
(Here and hereinafter, the first errors are statistical and the second are systematic.) The $\chi_{c2}(3930)$ state is a 
good candidate of $\chi_{c2}(2P)$, but this would imply that the hyperfine splitting of $12~\mevcs$ between $\chicpt$ 
and $X(3915)$ would be only 6\% of that between $\chi_{c2}(1P)$ and $\chi_{c0}(1P)$~\cite{guofk}. In contrast, an early 
calculation~\cite{theo} utilizing the Godfrey-Isgur relativistic potential model~\cite{GI_model} predicts a much larger 
mass difference of about $60~\mevcs$~\cite{theo}. The $X^*(3860)$ observed by Belle is another candidate of 
$\chi_{c0}(2P)$ but it was not seen in LHCb's study of the $B^+\to K^+D^+D^-$ decay~\cite{ddb-lhcb}. One interpretation 
of the $X^*(3860)$ is a $\ddb$ bound state close to the threshold with isospin $I=0$~\cite{wang_e}. Therefore, 
additional studies of the $P$-wave triplet states near $3.9~\gevcs$ are needed for a more comprehensive understanding 
of the $XYZ$ states and, in particular, of the $X(3872)$.

Both $0^{++}$ and $2^{++}$ states can be produced in two-photon collisions and can decay to $\gpsp$ via an E1 
transition. For example, the partial widths are expected to be $\Gamma(\chi_{c0}(2P)\to \gpsp) \approx 135~\kev$ and 
$\Gamma (\chi_{c2}(2P) \to \gpsp)\approx 207~\kev$ according to the aforementioned calculation~\cite{theo}. In this 
Article, we report an investigation of the $\gpsp$ final state produced in two-photon collisions ($\EE \to \EE \GG \to 
\EE \gpsp$ or $\GG\to\gpsp$ for brevity), using data collected with the Belle detector~\cite{Belle} at the KEKB 
asymmetric-energy $\EE$ collider~\cite{KEKB}. The $\psp$ is reconstructed from its hadronic final state $\pp\jpsi$ with 
$\jpsi$ reconstructed from a lepton pair $\LL~(\ell=e,\mu)$.  

\section{Detector, data sample, and Monte Carlo (MC) simulation}

The Belle detector is a large-solid-angle magnetic spectrometer that consists of a silicon vertex detector, a 50-layer 
central drift chamber, an array of aerogel threshold Cherenkov counters, a barrel-like arrangement of time-of-flight 
scintillation counters, and an electromagnetic calorimeter (ECL) comprised of CsI(Tl) crystals located inside a 
superconducting solenoid coil that provides a 1.5T magnetic field. An iron flux return located outside of the coil is 
instrumented to detect $K^0_{\rm L}$ mesons and to identify muons. The origin of the coordinate system is defined as the 
position of the nominal interaction point (IP). The $z$ axis is aligned with the direction opposite the $e^+$ beam and 
is parallel to the direction of the magnetic field within the solenoid. The $x$ axis is horizontal and points toward the 
 outside of the storage ring; the $y$ axis is vertical upward. The polar angle $\theta$ and azimuthal angle $\phi$ are 
measured relative to the positive $z$ and $x$ axes, respectively.

The integrated luminosity of Belle data used in this analysis is $980~\infb$. About 70\% of the data are collected at 
the $\yfos$ resonance, and the rest are taken at other $\yns$ ($n=1$, 2, 3, or 5) states or center-of-mass energies a 
few tens of $\mev$ below the $\Upsilon$ states. The {\sc Treps} event generator~\cite{TREPS} is used to simulate the 
signals of $\GG \to X\to \gpsp$ for optimization of selection criteria, efficiency determination and calculation of the 
luminosity function $L_{\GG}$ of two-photon collisions in Belle data. The major background is found to be the 
initial-state radiation (ISR) process $\EE\to\psp$, which has a cross section of $15.42\pm 0.12 \pm 0.89 ~\pb$ in the 
Belle data sample~\cite{belle_y4260}. There are $0.6\times 10^6$ events with a $\pp\LL$ final state in data, and an MC 
sample containing $3.8\times 10^6$ such events is simulated with the {\sc Phokhara} generator, which has a precision 
better than 0.5\%~\cite{phokhara}. An MC simulation using GEANT3~\cite{geant3} is used to model the performance of the 
Belle detector.

\section{Selection criteria and signal reconstructions}

Photon candidates are reconstructed from ECL clusters that do not match any charged tracks; the candidate with the 
highest energy is selected to form the $\gpsp$ final state. This energy is required to be larger than $100~\mev$ to 
suppress the background from fake photons. A candidate of $\psp\to\pp\jpsi$ with $\jpsi \to \EE$ or $\MM$ is 
reconstructed from four well-measured charged tracks, each having impact parameters with respect to the IP of 
$|dz|<5~\cm$ along the $z$ (positron-beam) axis and $dr < 0.5~\cm$ in the transverse $r$-$\phi$ plane. For a charged 
track, information from the detector subsystems is combined to form a likelihood $\mathcal{L}_i$ for a particle species 
of $i \in \{e, ~\mu, ~\pi, ~K ~\hbox{or proton}\}$~\cite{pid}. Tracks with $\displaystyle \mathcal{R}_K = 
\mathcal{L}_K/(\mathcal{L}_K + \mathcal{L}_\pi) < 0.4$ are identified as pions with an efficiency of about 95\%, while 
6\% of kaons misidentified as pions. Similar likelihood ratios are formed for electron and muon 
identification~\cite{EID,MUID}. Both lepton candidates are required to have $\mathcal{R}_e > 0.1$ for the $\jpsi\to\EE$ 
mode; at least one candidate is required to have $\mathcal{R}_{\mu} > 0.1$ for the $\jpsi\to \MM$ mode. For the first 
mode, any bremsstrahlung photons detected in the ECL within 0.05 radians of the original lepton direction are included 
in the calculation of the $\EE$ invariant mass. 

\begin{figure}[tbp]
\centering
 \psfig{file=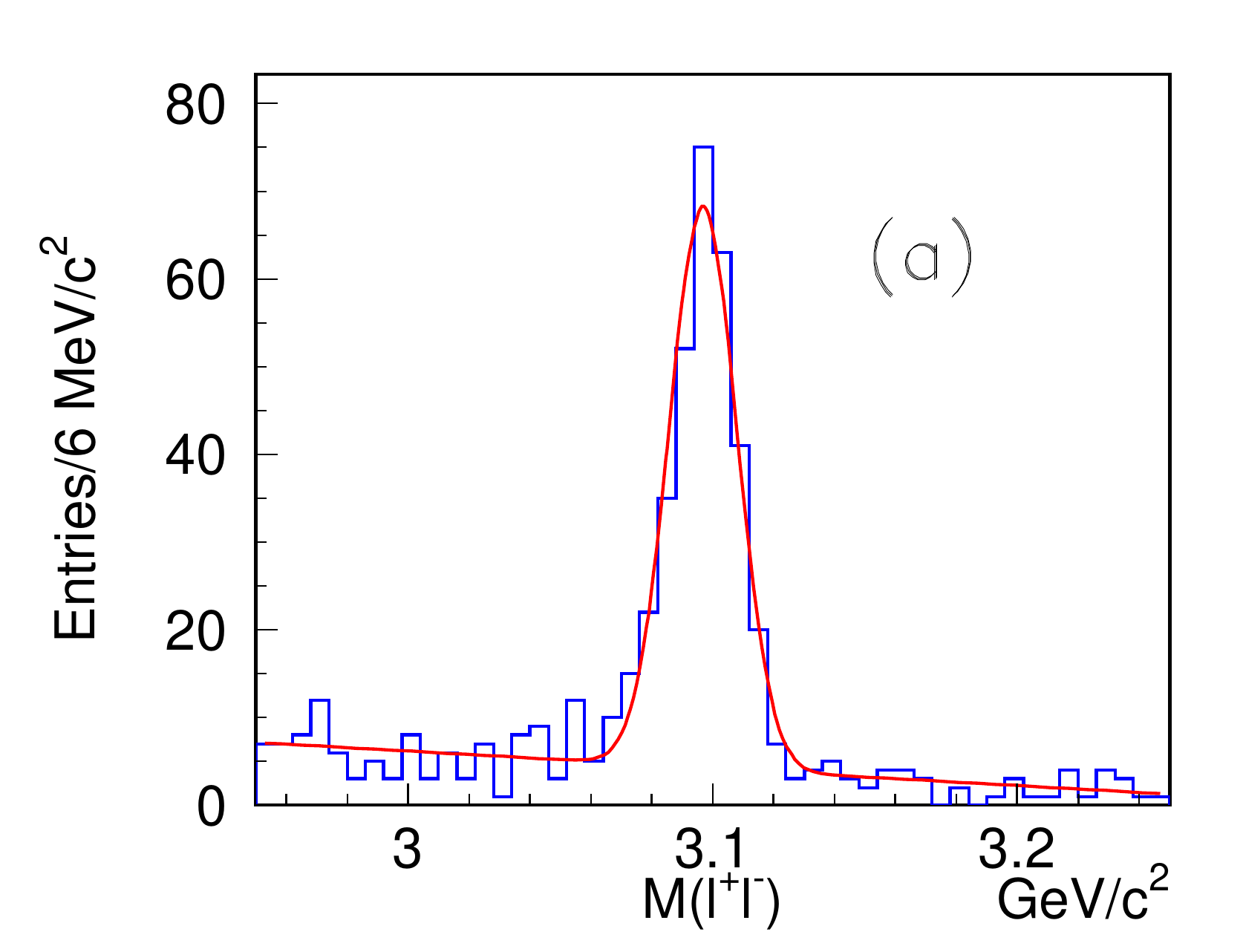,width=.45\textwidth}
\psfig{file=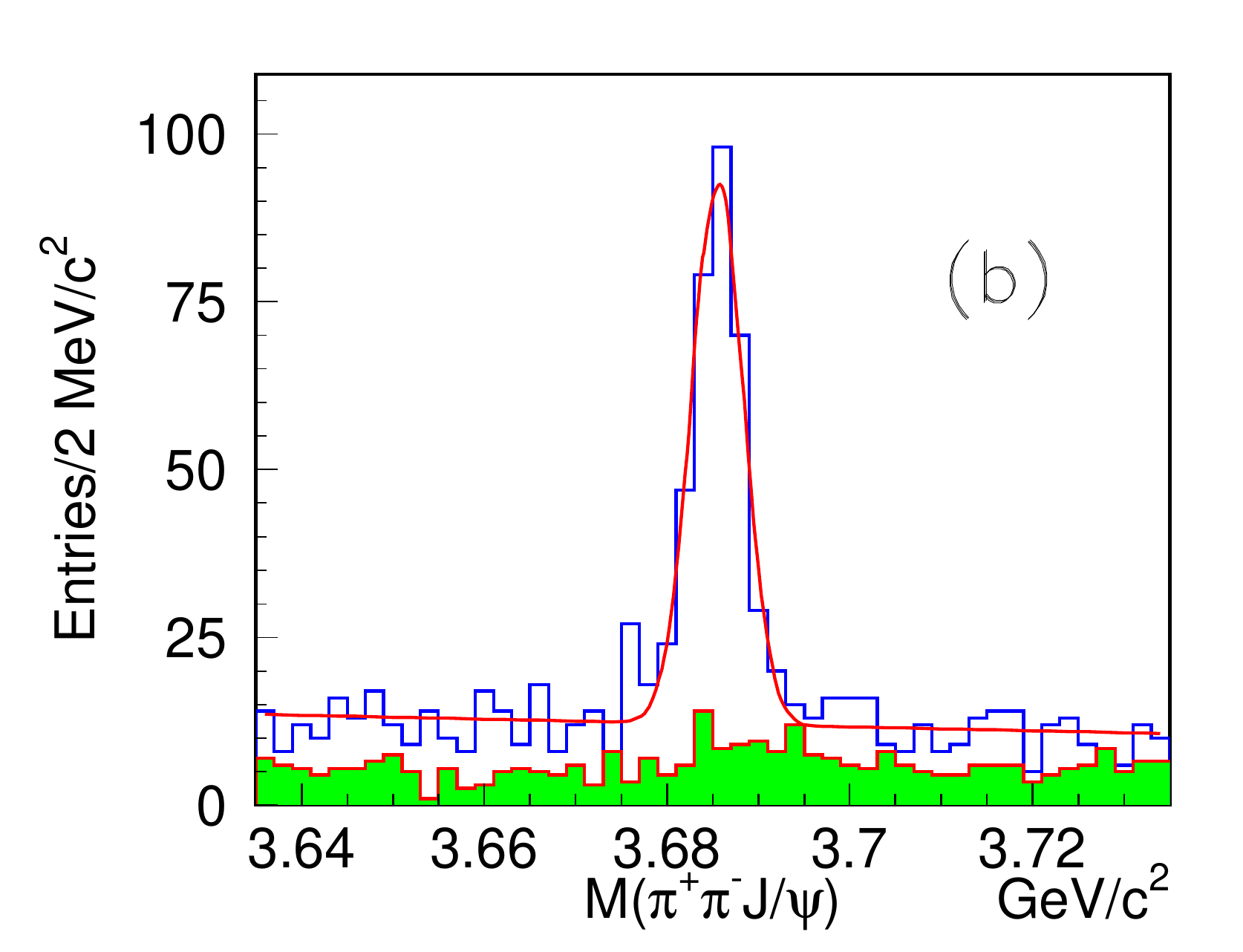,width=.45\textwidth}
\caption{
Invariant mass distributions of (a) $\LL(l=e,\mu)$ for the $\jpsi$ signal and (b) $\pp\jpsi$ for the $\psp$ signal 
in data. The shaded histogram is the background estimated from the $\jpsi$ mass sidebands. The curves show the 
best fit results. }
\label{mll-mppjpsi}
\end{figure}

The invariant mass distributions of the lepton pair ($M_{\LL}$) from data are shown in Fig.~\ref{mll-mppjpsi}(a), where 
clear $\jpsi$ signals are seen. By fitting the $M_{\LL}$ distributions with a Gaussian function for the $\jpsi$ signal 
and a first-order polynomial function for background, we obtain the $\jpsi$ mass resolutions of $\sigma_{\LL}^{\rm 
data} = 11.0\pm 0.6~\mevcs$ from data and $\sigma_{\LL}^{\rm MC} = 9.4\pm 0.1~\mevcs$ from signal MC simulation. A 
lepton pair is regarded as a $\jpsi$ candidate if $|M_{\LL}-m_{\jpsi}| < 4\sigma_{\LL}$, where $m_{\jpsi}$ is the 
nominal mass of $\jpsi$~\cite{PDG} and $\sigma_{\LL} = 11.0~\mevcs$. Events in the $\jpsi$ mass sidebands, defined as 
$M_{\LL} \in [m_{\jpsi}-13\sigma_{\LL}, m_{\jpsi}-5\sigma_{\LL}] \cup [ m_{\jpsi}+5\sigma_{\LL}, 
m_{\jpsi}+13\sigma_{\LL} ]$, are used to study the background in the $\psp$ reconstruction. Figure~\ref{mll-mppjpsi}(b) 
shows the distributions of $M_{\ppjpsi} \equiv M_{\pp\LL} - M_{\LL}+m_{\jpsi}$ from data, where $M_{\pp\LL}$ is the 
invariant mass of $\pp\LL$. Fitting the $M_{\ppjpsi}$ distributions with a Gaussian function for $\psp$ signal and a 
second-order polynomial function for the background, we obtain the $\psp$ mass resolutions of $\sigma_{\ppjpsi}^{\rm 
data} = 2.80\pm 0.21~\mevcs$ from data and $\sigma_{\ppjpsi}^{\rm MC} = 2.52\pm 0.04~\mevcs$ from the signal MC 
simulation. The $\psp$ signal window is defined to be $|M_{\ppjpsi} - m_{\psp}| < 2.5\sigma_{\ppjpsi}$, where $m_{\psp}$ 
is the nominal mass of $\psp$~\cite{PDG} and $\sigma_{\ppjpsi} = 2.80~\mevcs$. To estimate the background in the $\psp$ 
reconstruction, the sideband region is defined to be $|M_{\ppjpsi} - m_{\psp} \pm 9\sigma_{\ppjpsi}| < 
3.75\sigma_{\ppjpsi}$.

\begin{figure}[tbp] 
\centering
\includegraphics[width=.45\textwidth]{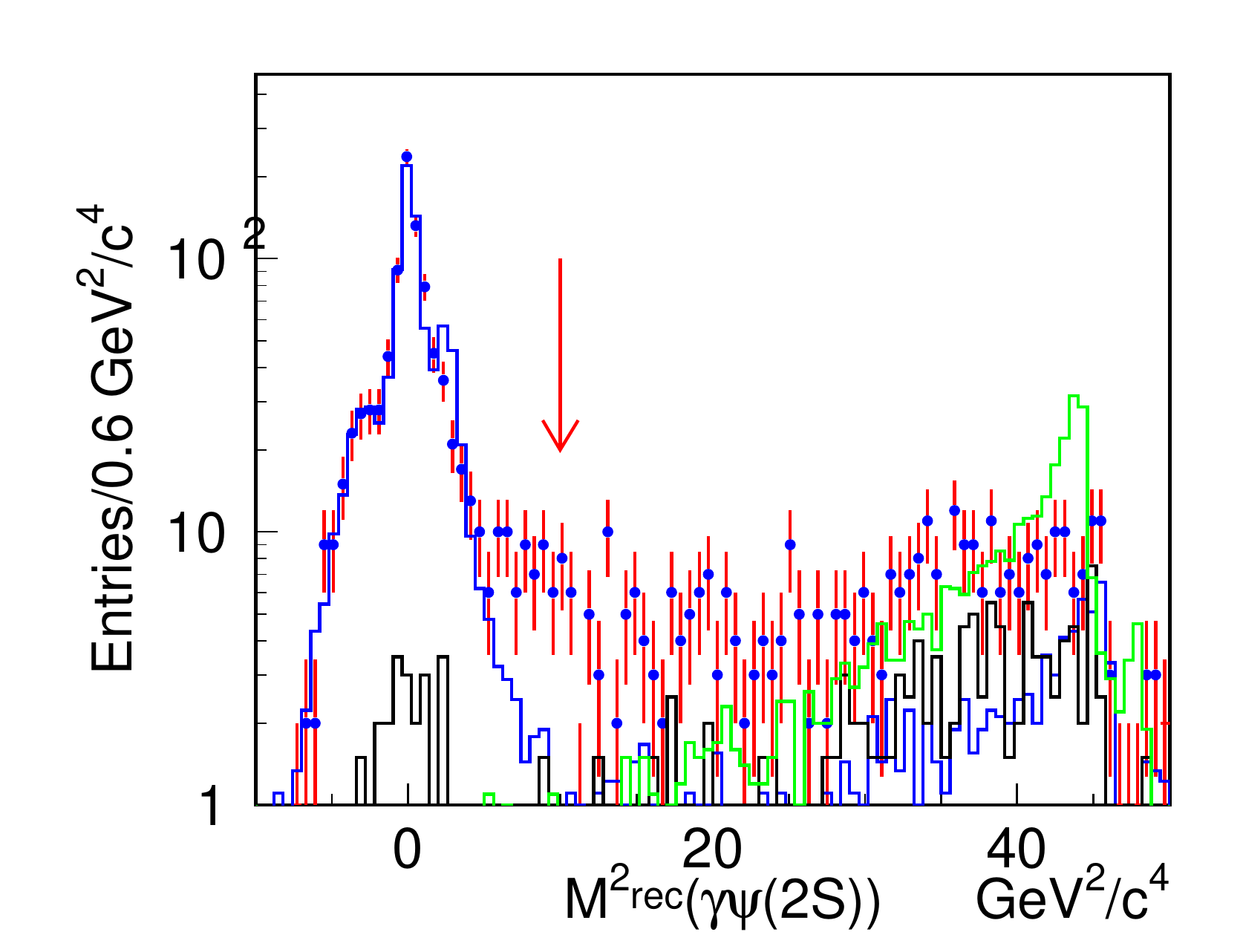}
\caption{The distributions of recoil mass square of $\gpsp$. The dots with error bars are data, the black blank 
histogram is the background estimated from the $\psp$ mass sidebands, the blue histogram is the ISR MC simulation and 
the green histogram is the signal MC simulation. The normalization of ISR MC simulation is according the region 
$\MMS(\gpsp) < 10 ~(\gevcs)^2$, and the one of the signal MC simulation is arbitrary. The arrow shows the position of 
$\MMS(\gpsp) = 10~(\gevcs)^2$. }
\label{mms} 
\end{figure}

The background is dominated by  $\EE\to\psp$ via ISR, where $\psp$ is combined with a fake photon. Figure~\ref{mms} 
shows the distributions of the recoil mass squared $\MMS(\gpsp)$ of $\gpsp$. For two-photon collision events, there may 
be an outgoing $\EE$ pair traveling back-to-back along the $\EE$ beams so that $\MMS(\gpsp)$, corresponding to the mass 
squared of the outgoing $\EE$ pair, tends to be large. For ISR events, the recoil of $\gpsp$ is dominated by one 
energetic ISR photon with $E(\gisr) > 1.5~\gev$, so $\MMS(\gpsp)$ is around zero. We apply $\MMS(\gpsp)> 10~(\gevcs)^2$ 
to remove most ISR events. Nevertheless, there still remain events with two ISR photons traveling back-to-back along the 
$\EE$ collision beams; such events have a topology similar to two-photon collisions. 

\begin{figure}[tbp] 
\centering
\psfig{file=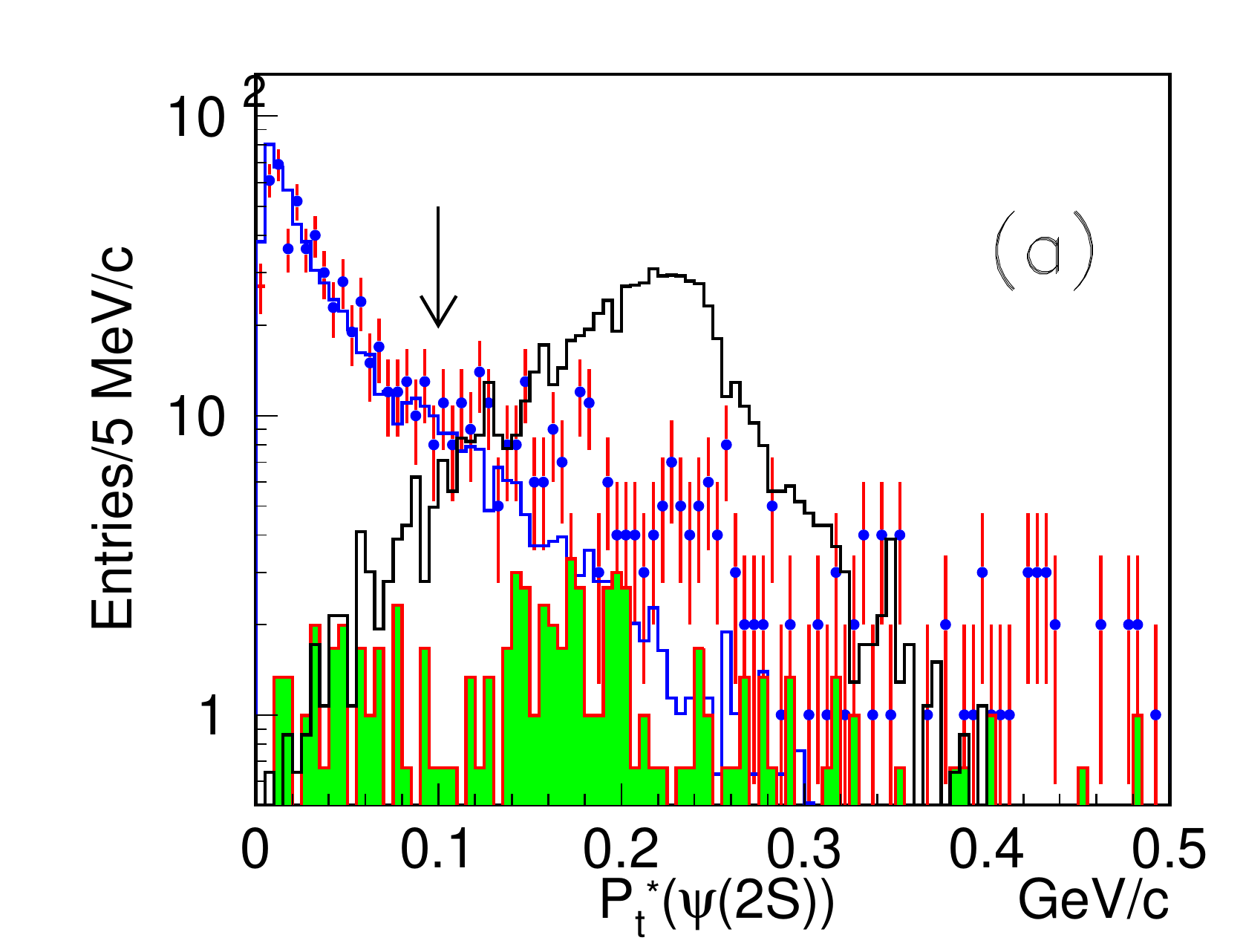, height=5.0cm}
\psfig{file=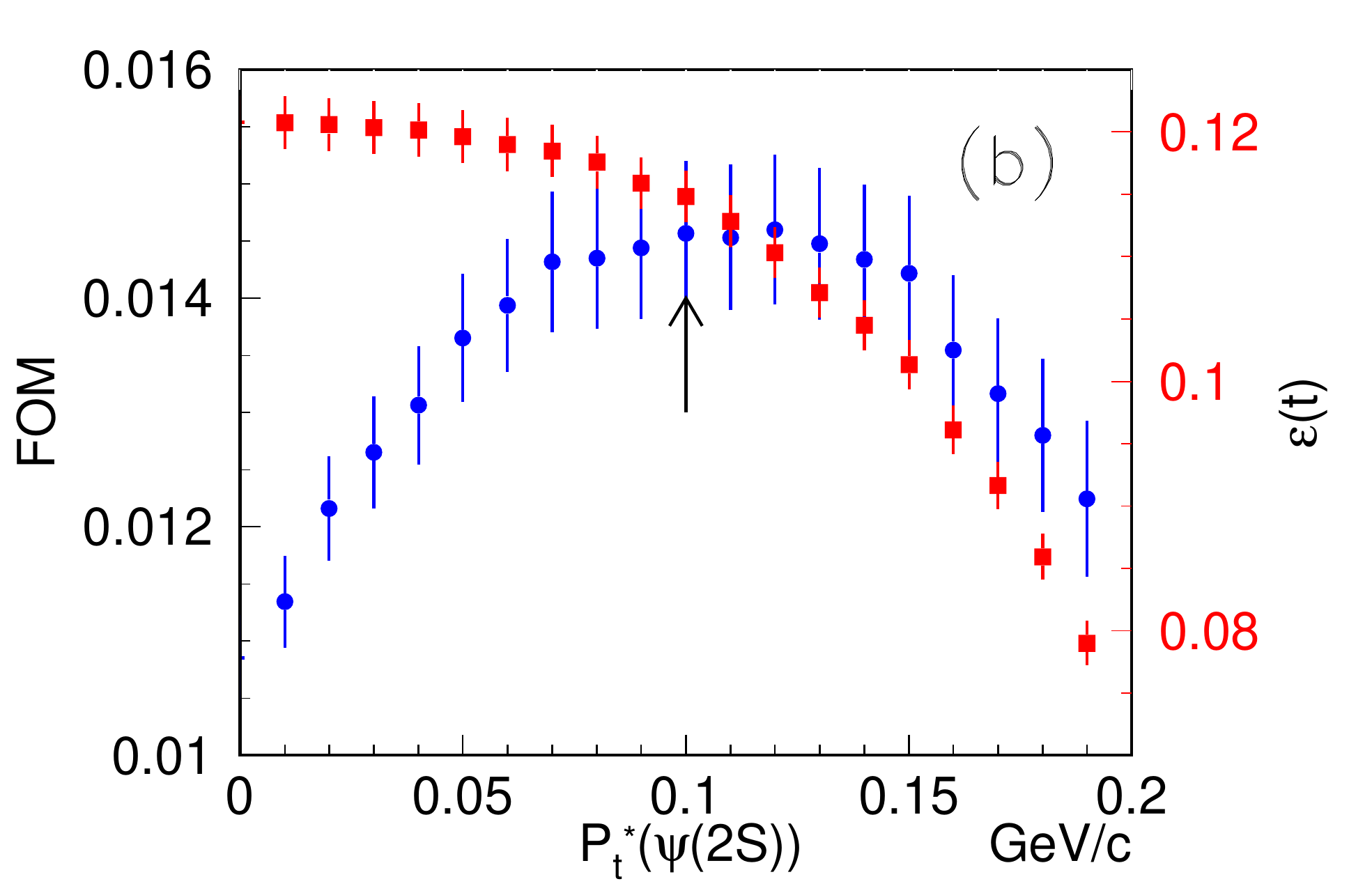, height=5.0cm}\\
\psfig{file=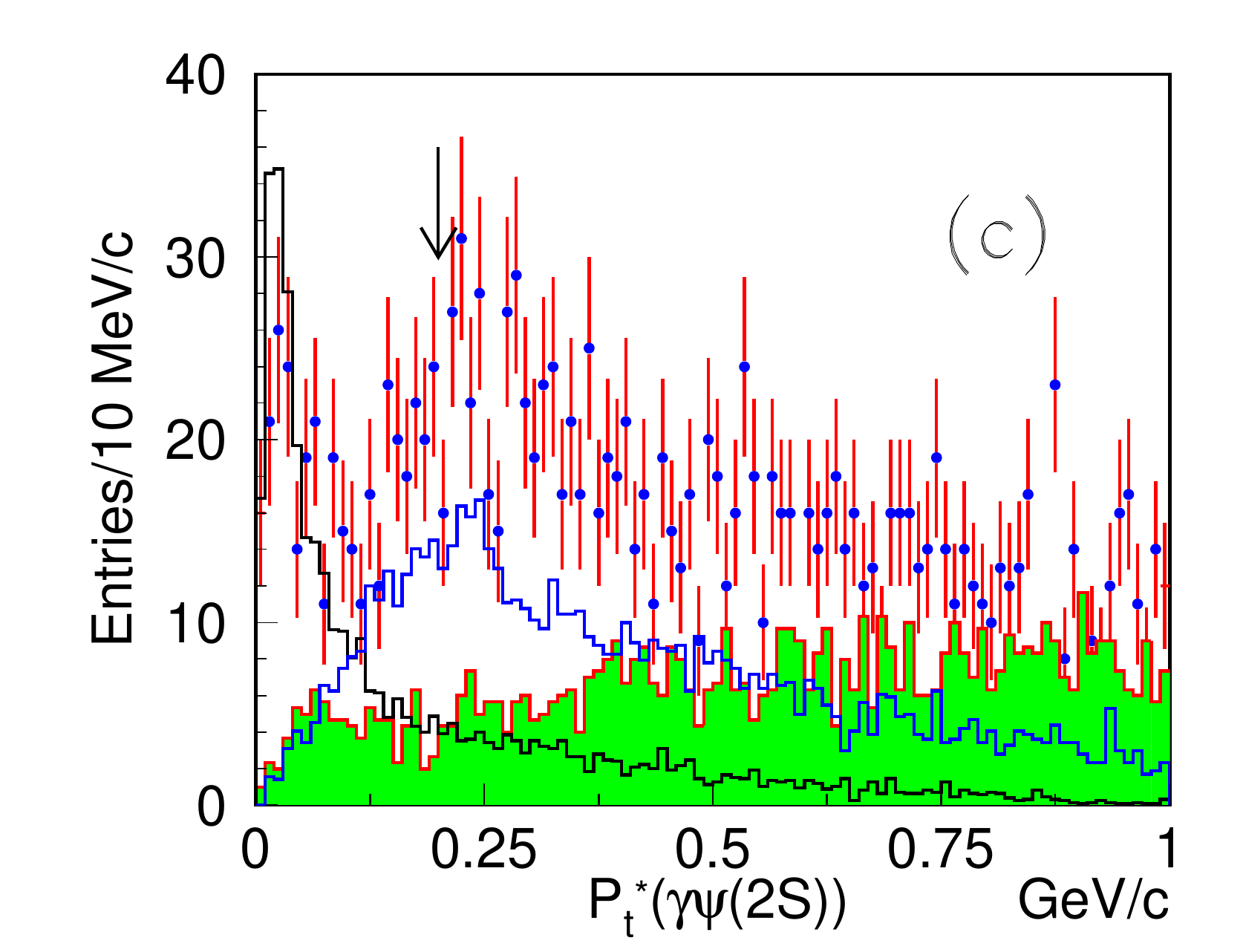, height=5.0cm}
\psfig{file=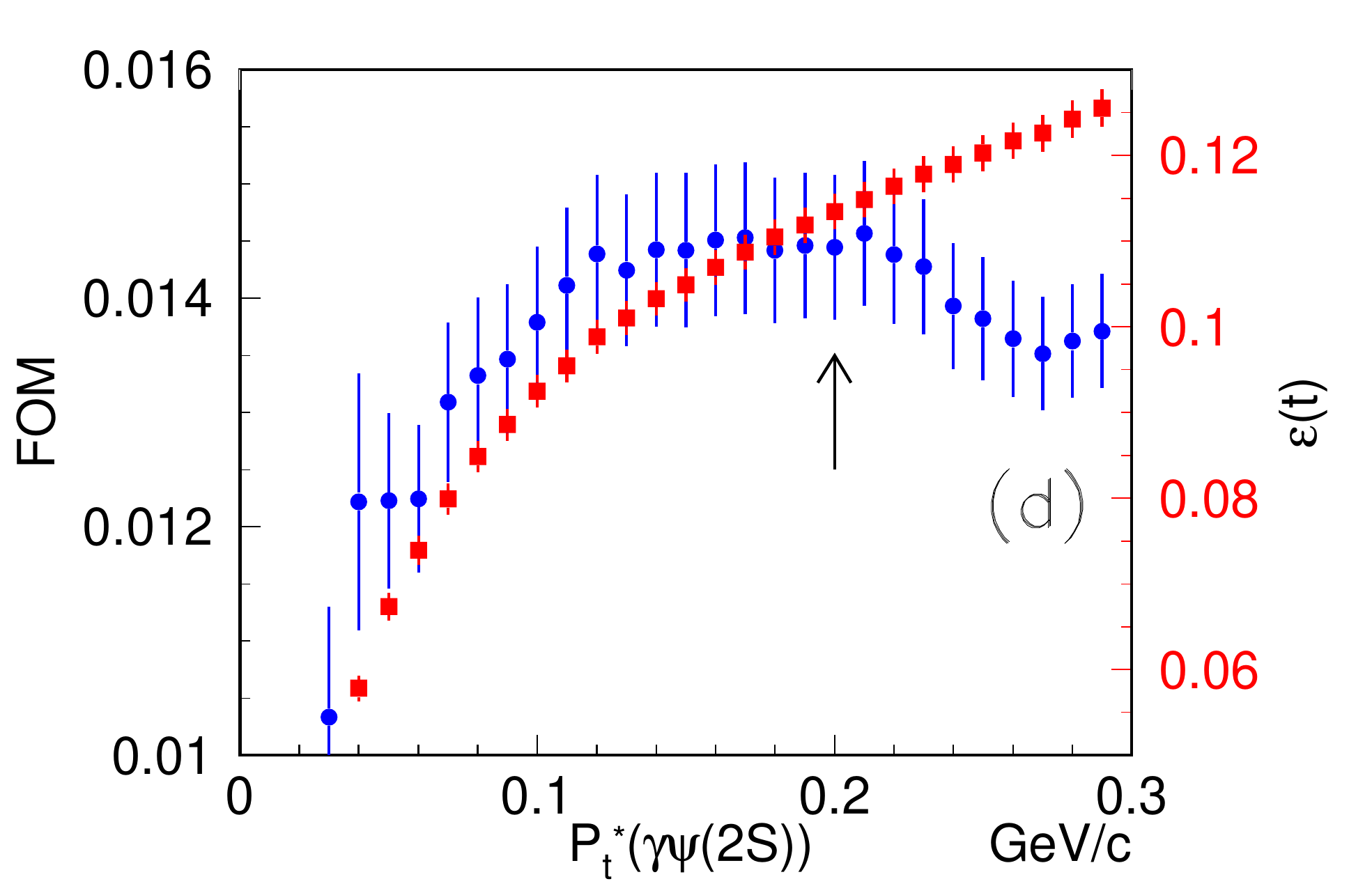, height=5.0cm}
\caption{ 
The distributions of transverse momenta of $\psp$ (top row) and $\gpsp$ (bottom row) in the c.m.s of $\EE$ collision. In 
the left panel, the dots with error bars are from the signal region, the shaded histograms are the backgrounds estimated 
from the $\psp$ mass sidebands, the black blank histograms are signal MC simulation with arbitrary normalizations, and 
the blue blank histograms are ISR MC simulation. The normalization of ISR MC simulation in (a) is according to data with 
$P_t^*(\psp) < 0.1~\gevcs$, and the one in (c) is according to the size of ISR MC simulation sample. The right panel 
shows the Punzi FOMs in dots and efficiencies in diamonds versus (b) $P_t^*(\psp)$ and (d) $P_t^*(\gpsp)$. The arrows 
show the selections applied. } 
\label{pt_tot} 
\end{figure}

To suppress the ISR background further, the transverse momenta of $\psp$ and $\gpsp$, i.e., $P^*_t(\psp)$ and 
$P^*_t(\gpsp)$, calculated in the center-of-mass system (c.m.s.) and shown in Fig.~\ref{pt_tot} (a) and (c), are used. 
$P^*_t(\gpsp)$ is small for most of the signal events, in which the outgoing $\EE$ travel along the accelerator 
beamline. However, $P^*_t(\psp)$ could be large if $\psp$ originates from the decay of a resonance such as 
$\chi_{c0}(2P)$ or $\chi_{c2}(2P)$. For the ISR events, $P_t^*(\psp)$ is small since the ISR photon(s) always travel 
along the accelerator beamline. We optimize the selections of $P^*_t(\psp)$ and $P^*_t(\gpsp)$ based on the Punzi 
figure of merit (FOM), defined as
\beq
{\rm FOM}  \equiv \frac{\eff(t)}{a/2+\sqrt{N_{\rm bkg}(t)}} 
\eeq 
according to Eq.(7) in Ref.~\cite{Punzi}. Here, $\eff(t)$ is the signal efficiency based on the selection criterion $t$, 
$a$ is the number of sigmas corresponding to one-side Gaussian tests --- we take $a=5$ --- and $N_{\rm bkg}(t)$ is the 
background estimated from the ISR events and the $\psp$ mass sidebands. The FOM and $\eff(t)$ versus $P_t^*$ selections 
are shown in Figs.~\ref{pt_tot}(b) and \ref{pt_tot}(d). We apply $P^*_t(\psp) > 0.1~\gevc$ and $P^*_t(\gamma\psp) < 
0.2~\gevc$ with selection efficiencies of $\eff^{\rm MC}(t) = (97.1\pm 0.3)\%$ and $\eff^{\rm MC}(t) = (67.8\pm 
0.7)~\%$, respectively. There are about 150 ISR events surviving these selection criteria with an efficiency of about 
$0.02\%$.

\section{Invariant mass distribution of $\gpsp$ and two structures}
\label{sec5}

\begin{figure}[tbp]
\centering
 \psfig{file=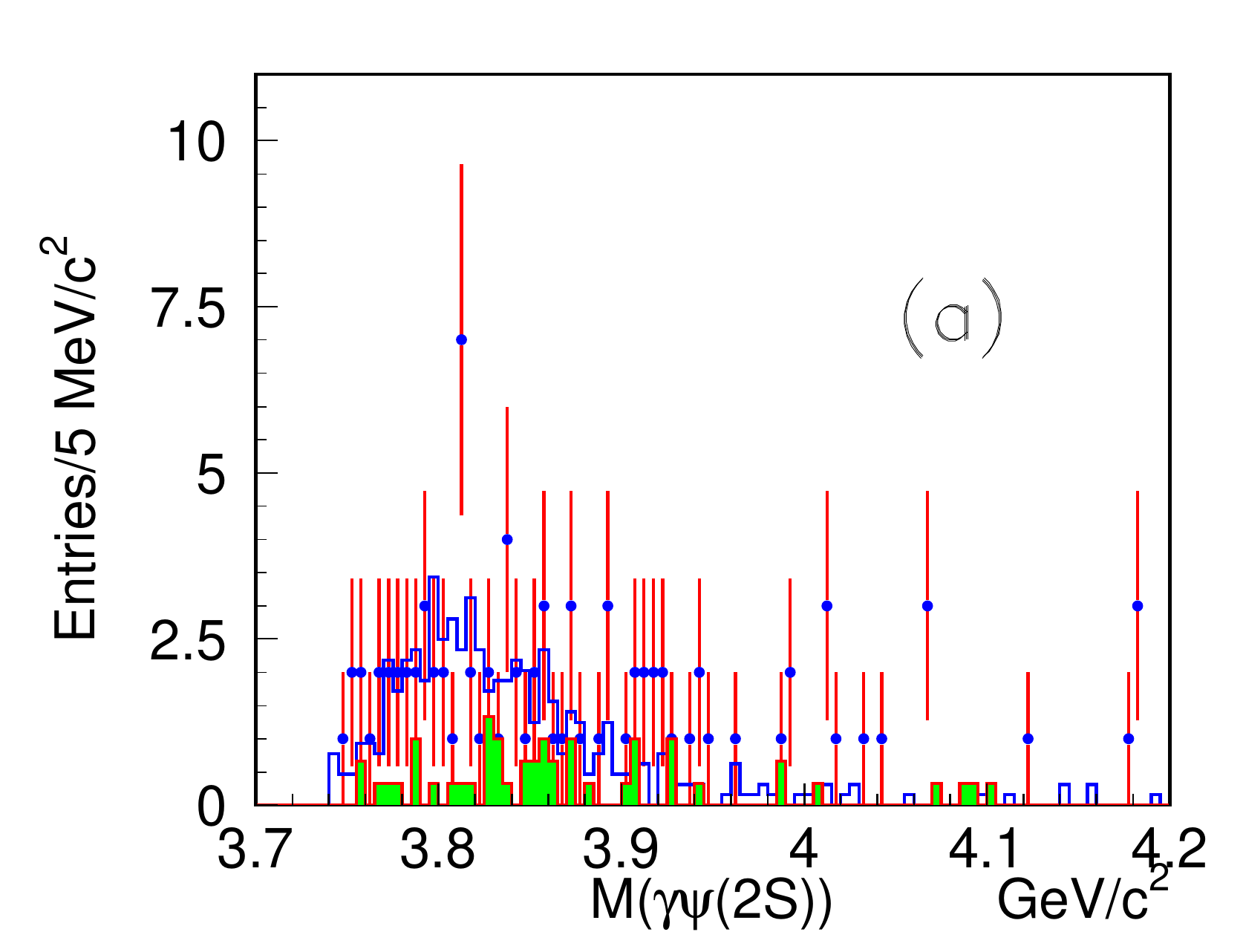,width=.45\textwidth}
 \psfig{file=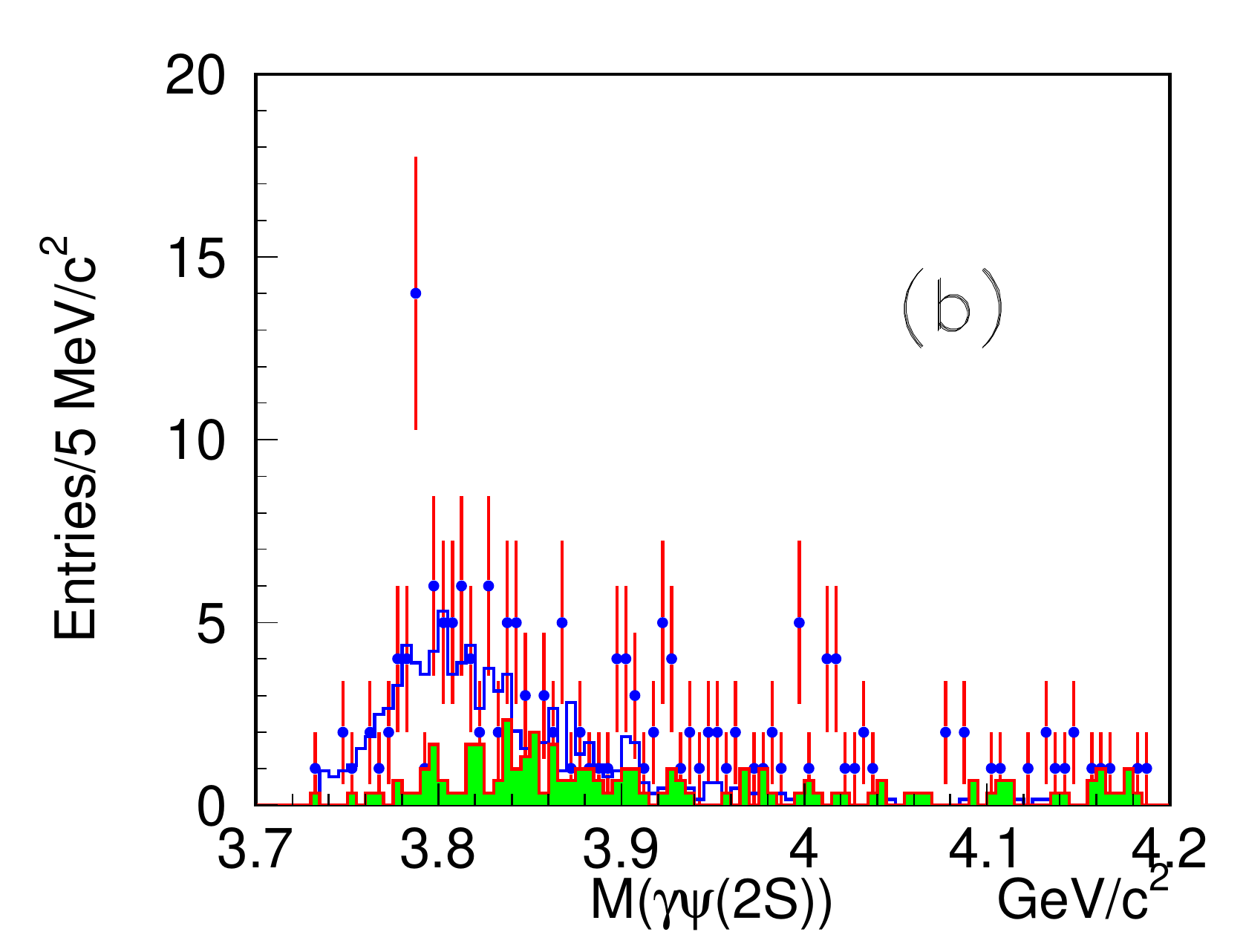,width=.45\textwidth}
\caption{The invariant mass distributions of $\gpsp$ from (a) $\EE$ mode and (b) $\MM$ mode. The dots with error bars 
are data, and the shaded histograms are backgrounds estimated from the $\psp$ mass sidebands. The blank histograms are 
ISR events simulated and scaled to the size of the Belle data sample.}
\label{mgpsp}
\end{figure}

\begin{figure}[tbp] 
\centering
 \psfig{file=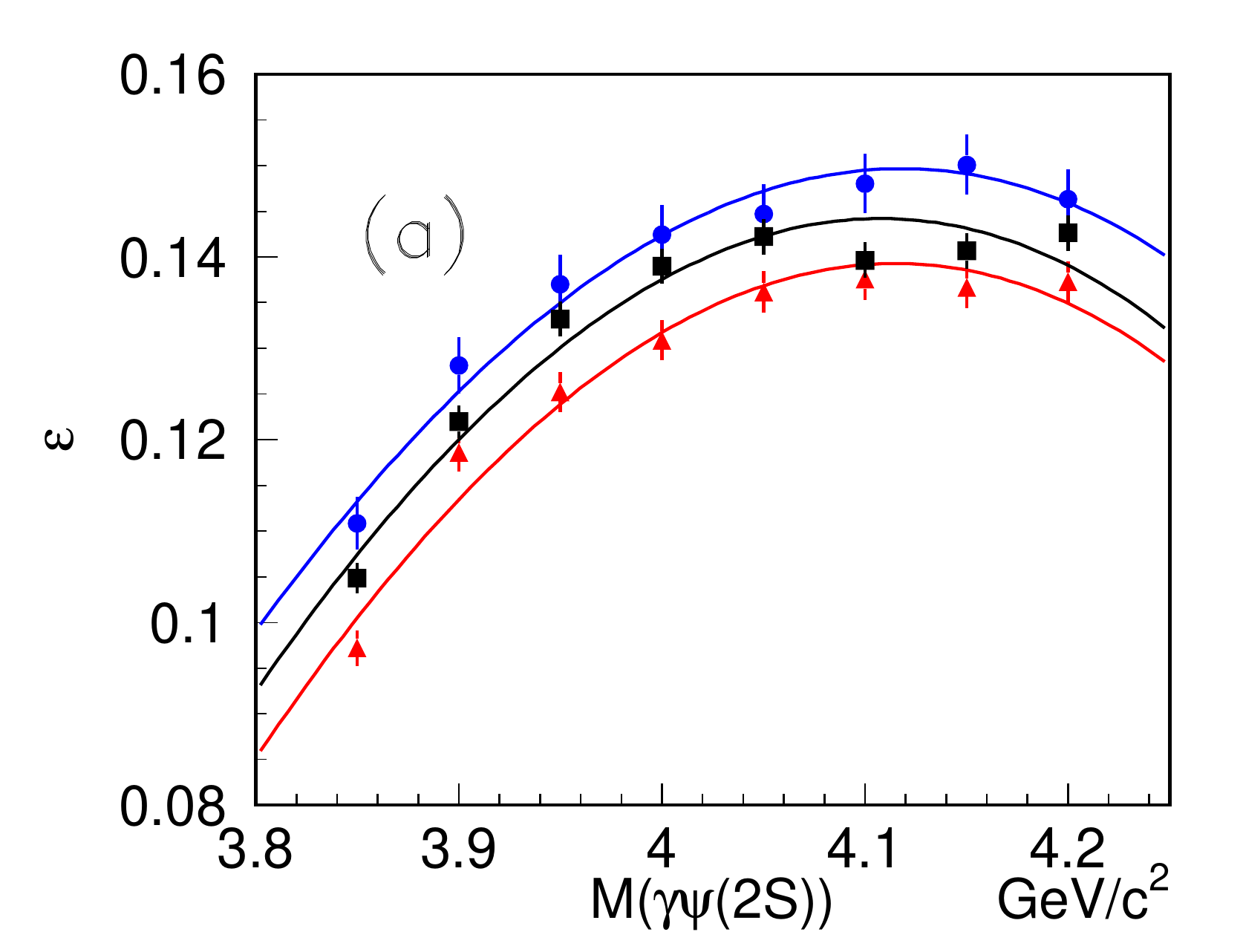,width=.45\textwidth}
 \psfig{file=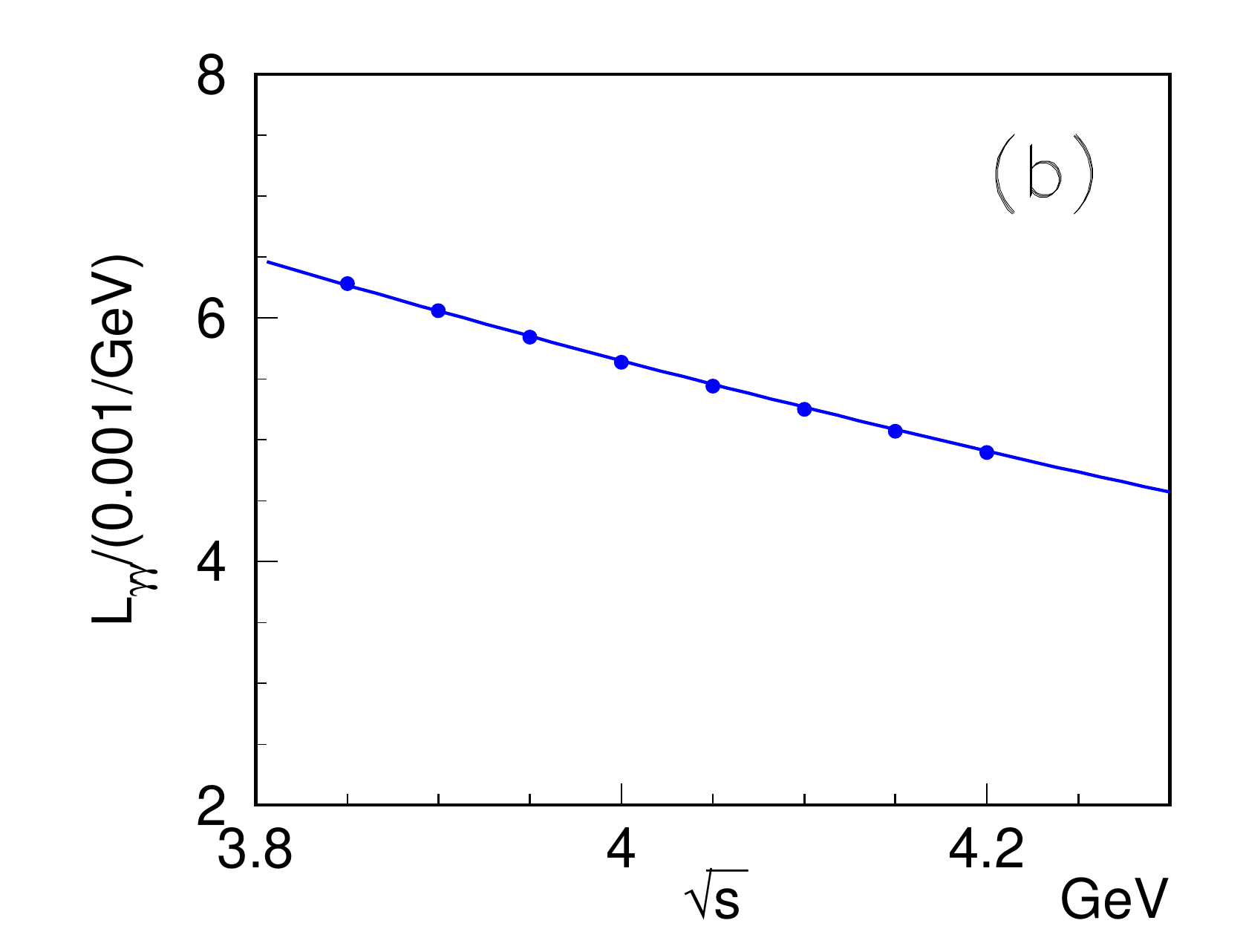,width=.45\textwidth}
\caption{(a) The efficiencies at different $M_{\gpsp}$ from MC simulation, and (b) the two-photon luminosity function 
$L_{\GG}(\sqrt{s})$. In (a), the blue dots are the efficiencies for $\jpc = 0^{++}$, and the black (red) dots are the 
efficiencies for $2^{++}$ with helicity $|\lambda| = 0$ ($|\lambda|=2$). The curves show the best fits with polynomial 
functions.} 
\label{eff_lum} 
\end{figure}

Figure~\ref{mgpsp} shows the invariant mass distributions for data of $\gpsp$ ($M_{\gpsp}$) in the $\jpsi\to\EE$ and 
$\MM$ modes. The distributions of the backgrounds estimated from the scaled $\psp$ mass sidebands and ISR events 
simulated by {\sc Phokhara}~\cite{phokhara} are also shown in Fig.~\ref{mgpsp}. The ratio between data and ISR MC 
simulation is $0.147\pm 0.012$ from the distributions in the region $ M_{\gpsp}< 3.9~\gevcs$,  while the expected ratio 
is $0.156\pm 0.009$ according to the cross section and the size of the ISR MC sample. Figure~\ref{eff_lum} shows the 
signal selection efficiency and the two-photon luminosity function $L_{\GG}(\sqrt{s})$, which is defined as the 
probability of a two-photon emission with $\GG$ c.m.s. energy $\sqrt{s}$ in the Belle experiment~\cite{TREPS}. The 
efficiencies for $\jpc = 0^{++}$ and $2^{++}$ ($|\lambda|=0$ or 2) range from $\sim 10\%$ to $\sim 15\%$ for $M_{\gpsp}$ 
between $3.85~\gevcs$ and $4.20~\gevcs$. The difference of $L_{\GG}$ between $0^{++}$ and $2^{++}$ is very small and 
thus ignored.

\begin{figure}[tbp]
\centering
\psfig{file=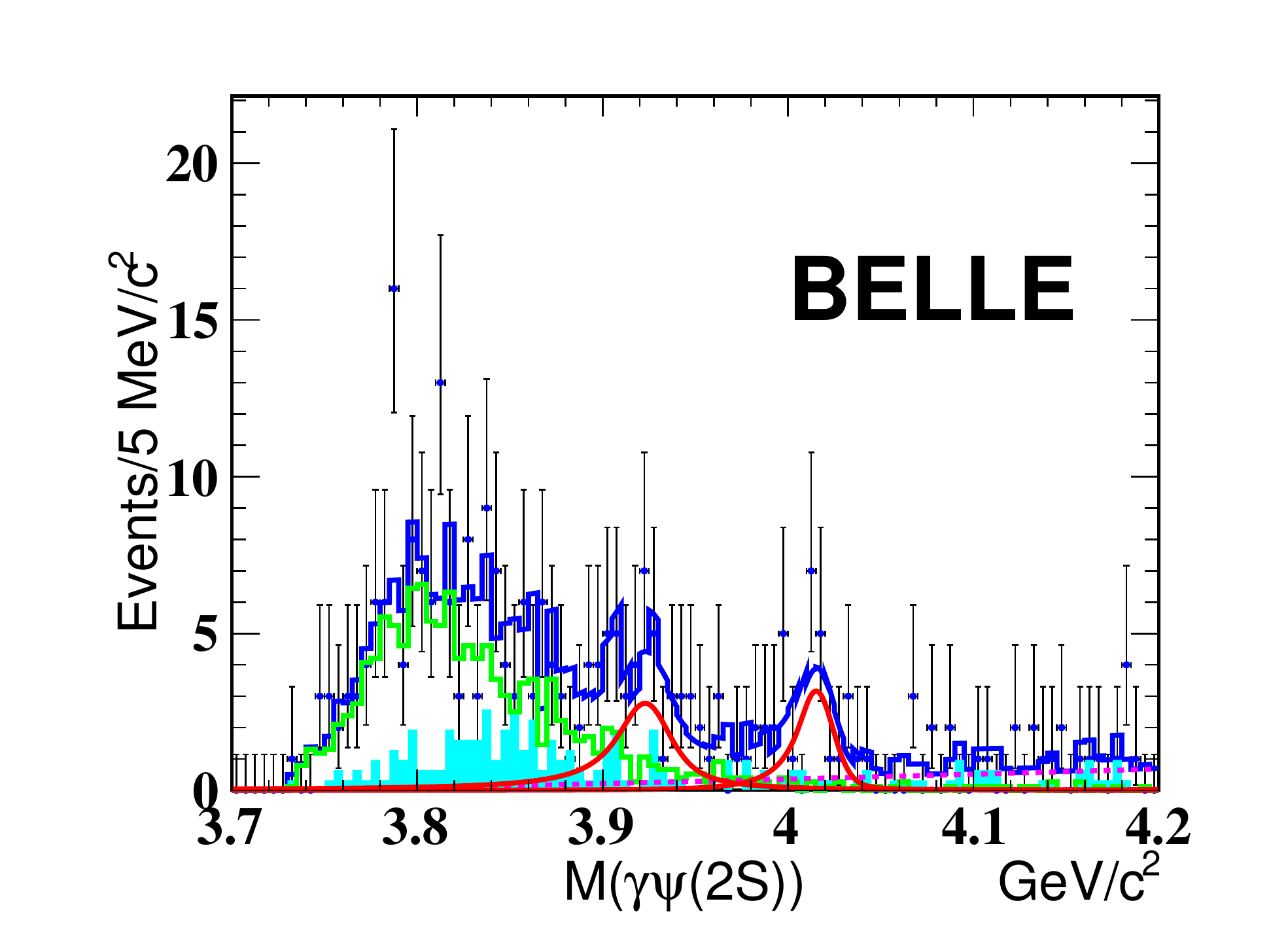, width=0.60\textwidth}
\caption{The $\gpsp$ invariant mass distribution and the fit result. The points with error bars show the data while the 
shaded histogram is the normalized background from the $\psp$ mass sidebands. The solid blue curve shows the best fit 
results. The red signal curves from the convolutions of BW and CB functions show the contributions from the two 
structures. The green blank histogram shows the component of ISR events of $\EE\to\psp\to\pp\jpsi$. The pink dashed 
line shows the possible additional background, modeled by a second-order polynomial.} 
\label{fit}
\end{figure}

The $M_{\gpsp}$ distribution of data after combining the $\EE$ and $\MM$ modes are shown in Fig.~\ref{fit}. Excesses 
around $3.92~\gevcs$ and $4.02~\gevcs$ are seen. To study the excesses, a binned extended maximum-likelihood fit is 
performed to the $M_{\gpsp}$ mass spectra. The function used for the fit is characterized by the sum:
\beq\label{full_pdf}
f_{\rm sum} = f_{\rm R_1} + f_{\rm R_2} + f_{\rm ISR} + f_{\rm bkg} +f_{\rm SB}.
\eeq
Here, $f_{\rm R_1}$ ($f_{\rm R_2}$) is for the structure $R_1$ ($R_2$) near $3.92~\gevcs$ ($4.02~\gevcs$), $f_{\rm ISR}$ 
for the ISR events, $f_{\rm SB}$ for the background in $\psp$ reconstruction, and $f_{\rm bkg}$ for the possible 
additional backgrounds. The $f_{\rm SB}$ distribution is estimated from the $\psp$ mass sidebands, and its yield is 
fixed in the fits. The function $f_{\rm R_1}$ ($f_{\rm R_2}$) contains the convolution of a relativistic Breit-Wigner 
(BW) function with a form of $ 12\pi \Gamma_{\GG} \Gamma_X/((s - M^2)^2 + M^2\Gamma^2)$ and a Crystal Ball (CB) 
function~\cite{CB_func} with a mass resolution of about $7.2~\mevcs$ ($8.0~\mevcs$), and the parameters of CB function 
are fixed according to the signal MC simulation of a resonance with a mass near the one of $R_1$ ($R_2$) states and with 
no width. The resonant parameters in the BW function, $M$, $\Gamma$, and $\Gamma_X\Gamma_{\GG}$ are the mass, the width, 
and the product of partial widths of the decays to the final state $X$ ($= \gpsp$ here) and $\GG$, respectively. The 
product $\Gamma_X\Gamma_{\GG}$ is treated as one parameter, since it is impossible to separate $\Gamma_X$ and 
$\Gamma_{\GG}$ in the fits. The efficiency curve $\eff$ is shown in Fig.~\ref{eff_lum}(a) and is incorporated into 
$f_{\rm R_1}$ and $f_{\rm R_2}$, i.e., $f_{\rm R} \propto \eff\cdot (\rm BW\otimes CB)$. The widths $\Gamma_{R_1}$ and 
$\Gamma_{R_2}$ are found to be small and thus the possible interference between $R_1$ and $R_2$ is expected to be small 
and is ignored in the fit. The histogram of $M_{\gpsp}$ distribution from the ISR MC simulation is used for $f_{\rm 
ISR}$. There may be more subdominant sources of background, such as high order QED processes and continuum production of 
$\GG\to\gpsp$, but their individual and collective contributions are not clearly distinguishable with the current 
limited statistics. A second-order polynomial function is used for $f_{\rm bkg}$, and polynomial functions with 
different order are considered to estimate the systematic uncertainty. 

\begin{table}[tbp]
\centering
\begin{tabular}{| c | c | c | }
\hline
        Resonant parameters & $J = 0$ & $J=2$ \\\hline
        $M_{R_1}$ & \multicolumn{2}{c|}{$3922.4\pm 6.5 \pm 2.0$}    \\
        $\Gamma_{R_1}$ & \multicolumn{2}{c|}{$22\pm 17 \pm 4$}  \\
        $\Gamma_{\GG}\BR(R_1\to \gpsp)$ & $9.8\pm 3.6 \pm 1.2 $ & $2.0\pm 0.7\pm 0.2$ \\\hline
        $M_{R_2}$ & \multicolumn{2}{c|}{$4014.3\pm 4.0 \pm 1.5$}    \\
        $\Gamma_{R_2}$ & \multicolumn{2}{c|}{$4\pm 11 \pm 6$}  \\
        $\Gamma_{\GG}\BR(R_2\to \gpsp)$ & $6.2\pm 2.2\pm 0.8 $ & $1.2 \pm 0.4\pm 0.2$ \\\hline
\end{tabular}
\caption{Summary of the resonant parameters determined. The units of mass ($M$), width ($\Gamma$), product of partial 
width and branching fraction $\Gamma_{\GG}\BR$ are $\mevcs$, $\mev$ and $\ev$, respectively. The first  errors are 
statistical and the second are systematic.} 
\label{tab_summary}
\end{table}

The result from a fit in which all parameters are floated except the yield of the $f_{\rm SB}$ component is shown 
Fig.~\ref{fit} and Table~\ref{tab_summary}. The reduced chi-squared of the fit to the $M_{\gpsp}$ spectrum is 
$\chi^2/ndf = 0.69$. The signal yields are $N_{R_1} = 31\pm 11$ events for $R_1$ with $M_{R_1} = 3922.4\pm 6.5~\mevcs$ 
and $\Gamma_{R_1} = 22\pm 17~\mev$, and $N_{R_2} = 19\pm 7$ events for $R_2$ with $M_{R_2} = 4014.3\pm 4.0~\mevcs$ and 
$\Gamma_{R_2} = 4\pm 11~\mev$.   

The production of $R_1$ and $R_2$ in two-photon collisions are studied by determining the parameter 
$\BR\cdot\Gamma_{\GG}\equiv \Gamma_{\gpsp}\Gamma_{\GG}/\Gamma$ with the formula~\cite{TREPS}:
\beq\label{eq_3}
\BR\cdot\Gamma_{\GG} = \frac{n^{\rm sig}_{\rm fit}}{L_{\rm tot}\cdot \BR^{\rm prod}\cdot \eff\cdot F(\sqrt{s},J)}, 
\eeq
where $n^{\rm sig}_{\rm fit}$ is the signal yield from the fit, $L_{\rm tot} = 980~\infb$ is the integrated luminosity 
of the Belle data sample, $J$ is the spin of a structure, and $\BR^{\rm prod}$ is the product of branching fractions 
$\BR(\psp\to\ppjpsi) \cdot\BR(\jpsi\to\EE/\MM)$. Since $\Gamma_{R_1}$ and $\Gamma_{R_2}$ are small compared to the 
available kinetic energy in the decays, the spin-dependent factor is $F(\sqrt{s},J) = 4\pi^2(2J+1) L_{\GG}(\sqrt{s})/s$. 
The best fit gives $\Gamma_{\GG}\BR(R_1\to \gpsp) = (8.2\pm 3.4)~\ev$ if $J=0$ and $(1.6\pm 0.7)~\ev$ if $J=2$ for 
structure $R_1$, and $\Gamma_{\GG}\BR(R_2\to \gpsp) = (6.8\pm 2.8)~\ev$ if $J=0$ and $(1.4\pm 0.6)~\ev$ if $J=2$ for 
structure $R_2$. The ISR yield of $134 \pm 15$ is consistent with the estimate from the ISR MC simulation of $154\pm 
10$. The mass of $R_1$ indicates that it is a good candidate for $X(3915)$, $\chi_{c2}(3930)$ or an admixture of them. 
An alternate fit with both structures included and $M_{R_1}$ and $\Gamma_{R_1}$ fixed to the nominal $X(3915)$ 
parameters yields $\Gamma_{\GG}\BR(X(3915)\to\gpsp) = 9.6\pm 2.9 \pm 1.0~\ev$ if $\jpc = 0^{++}$ of $1.9\pm 0.6~\ev$ if 
$\jpc = 2^{++}$. Another alternative fit with the mass and width of $R_1$ fixed to those of $\chi_{c2}(3930)$ yields 
$\Gamma_{\GG} \BR(\chi_{c2}(3930)\to\gpsp) = 2.2\pm 0.6 \pm 0.4~\ev$ if $\jpc=2^{++}$. A third alternate fit with $R_1$ 
being an admixture of $X(3915)$ and $\chi_{c2}(3930)$ shows no notable change in the fit quality. 

The local signal significance is determined to be $3.5\sigma$ for $R_1$ and $3.4\sigma$ for $R_2$ by comparing the value 
of $\Delta (-2\ln \mathcal{L}) = -2\ln(\mathcal{L}_{\rm max}/\mathcal{L}_0)$ and the change of the number of free 
parameters ($N_{\rm par}$) in the fits, where $\mathcal{L}_{\rm max}$ is the likelihood with both $R_1$ and $R_2$ 
included in Eq.~\ref{full_pdf}, and $\mathcal{L}_0$ is the likelihood with only one of $R_1$ or $R_2$ excluded. The 
values of $-2\ln \mathcal{L}$, $\chi^2/ndf$, and $N_{\rm par}$ of all these fits are summarized in 
Table~\ref{tab_fit_quality}. Taking into account the systematic uncertainties, described in Section~\ref{sect_sys}, the 
lowest values of the local significance of $R_1$ is $3.1\sigma$. Since $R_2$ has never been seen before, the 
look-elsewhere effect is assessed for it with pseudo-experiments to check its global significance. The function for 
generating pseudo-experiments is $f_{\rm toyMC} = f_{\rm R_1}+f_{\rm ISR}+f_{\rm bkg}+f_{\rm SB}$ with the parameters 
from the nominal fit. The fit in each pseudo-experiment is performed with the same procedures as for the nominal fit to 
the actual data sample, except that the mass range of $R_2$ is limited to $M_{R_2}>3.95~\gevcs$ because the region 
$M_{\gpsp}<3.95~\gevcs$ is dominated by $R_1$ and ISR backgrounds. Among the $5.0\times 10^4$ pseudo-experiments, the 
number of experiments with $\Delta (-2\ln \mathcal{L})$ of $R_2$ signal larger than the one from data is 137. Therefore, 
the probability considering the look-elsewhere effect is about $(2.74\pm 0.23)\times 10^{-3}$, corresponding to a global 
significance of $2.8\sigma$. 

\begin{table}[htbp] 
\caption{The values of $-2\ln \mathcal{L}$, $\chi^2/ndf$ and number of free parameters ($N_{\rm par}$) in the different 
fits. From left to right, the rows are the fits with no resonance included, only $R_1$ included, only $R_2$ included, 
both $R_1$ and $R_2$ included (nominal fit), both resonances included and the mass and width of $R_1$ fixed to those of 
$X(3915)$, and both resonances included and the mass and width of $R_1$ fixed to those of $\chi_{c2}(3930)$. Only the 
differences among $-2\ln \mathcal{L}$ are meaningful in studying the statistical significance of $R_1$ and $R_2$.} 
\label{tab_fit_quality} 
\begin{tabular}{c | c | c | c | c | c | c } 
\hline 
--- & no resonance  & $R_1$ only & $R_2$ only & $R_1+R_2$ & $X(3915)+R_2$ & 
$\chi_{c2}(3930)+R_2$ \\\hline 
$-2\ln \mathcal{L}$ & $-2932.2$ & $-2946.5$ & $-2946.3$ & $-2965.4$ & $-2964.8$ & $-2963.0$ \\
$\chi^2/ndf$        & 0.76    & 0.74    & 0.78    & 0.70    & 0.68    & 0.68 \\
$N_{\rm par}$       & 5       & 8       & 8       & 11      & 9       &  9 \\\hline 
\end{tabular} 
\end{table}

\section{Systematic uncertainties}
\label{sect_sys}

There are systematic uncertainties in determining the resonant parameters of the two structures. The masses and widths 
are determined from fitting to the invariant mass distribution of $\gpsp$. In determining $\BR\cdot\Gamma_{\GG}$ with 
Eq.~(\ref{eq_3}), additional systematic uncertainties from the selection efficiency, the luminosity of Belle data sample 
and the branching fractions of $\jpsi$ and $\psp$ decays are taken into account.

The uncertainties due to the fits are estimated by changing the fit range, the number and the $f_{\rm SB}$ shape of 
the background in the $\psp$ reconstruction, the $f_{\rm bkg}$  shape, the bin width of the $M_{\gpsp}$ distribution, 
the parametrization of the BW function, and the resonant parameters of $X(3915)$ and $\chi_{c2}(3930)$. The fit range 
is changed from $[3.70, 4.20]~\gevcs$ to $[3.725,~4.15]~\gevcs$. The number of $\psp$ mass sideband events is changed 
by $1\sigma$, and the sideband region is changed from $|M_{\ppjpsi}-m_{\psp}\pm9\sigma_{\ppjpsi}|<3.75\sigma_{\ppjpsi}$ 
to $|M_{\ppjpsi} -m_{\psp}\pm 8\sigma_{\ppjpsi}|< 3.75\sigma_{\ppjpsi}$ to estimate the uncertainty due to the $f_{\rm 
SB}$ component. Another ISR MC sample is simulated to estimate the uncertainty from the shape of $f_{\rm ISR}$. The 
alternative polynomial function for $f_{\rm bkg}$ is first-order or  third-order. The bin width is changed from 
$5~\mevcs$ to $4~\mevcs$. The alternative formula of the resonant shape is $\displaystyle {\rm BW} \propto (M^2/s)\cdot 
12\pi\Gamma_{\GG} \Gamma_X/ ((s-M^2)^2+M^2\Gamma^2)$. The uncertainty from the resolution of $M_{\gpsp}$ is mainly 
related to the reconstructed $\gamma$, and it is estimated with a sample of about $4,000$ $\GG \to \chi_{c2} \to \gamma 
\jpsi$ events selected in the Belle data sample. Fitting to $\chi_{c2}$ signals in the $M_{\gamma\jpsi}$ distributions 
from data and MC simulation results in the consistent value of $10.84\pm 0.26~\mevcs$ and $10.77\pm 0.22~\mevcs$, 
respectively. Thus, the uncertainty due to the mass resolution of $\gpsp$ is expected to be very small and so is 
ignored. When $M_{R_1}$ and $\Gamma_{R_1}$ are fixed to those of $X(3915)$ or $\chi_{c2}(3930)$, their values are 
changed by $1\sigma$ to estimate the related systematic uncertainties~\cite{PDG}. The largest differences between the 
nominal fit results and those from these various fits are taken as the systematic uncertainties of the mass, the width 
and the product $\Gamma_{\GG}\BR(R\to \gpsp)$. 

Several sources of non-fit-related systematic uncertainties are considered. The particle identification uncertainty is 
2.8\%~\cite{pid,EID,MUID}; the uncertainty of the tracking efficiency is 0.35\% per track and is additive; the 
uncertainty of the photon reconstruction is 2\% per photon. The efficiency for the tracks in the extreme forward and 
backward regions obtained from MC simulation is found to be higher than that obtained in data according to the study of 
$\EE\to\psp\to\ppjpsi$ via ISR~\cite{belle_y4260}, and appropriate corrections have been applied. The uncertainty in the 
$\psp$ mass window requirement is measured to be 0.6\%, while the one of $\jpsi$ mass window is ignored. The 
efficiencies of the selection criteria on $P^*_t(\psp)$ and $P^*_t(\gpsp)$ are strongly related to the boost 
transformation from the lab system to the c.m.s. of $\EE$ collisions. However, the related uncertainty is very small, 
and 1\% is taken to be a conservative estimation for the uncertainty due to the $P_t^*$ selections. The uncertainty due 
to the $\MMS(\gpsp)$ requirement is less than 0.5\%. The uncertainty due to the momentum and angular distributions of 
helicities 0 and 2 for $J=2$ is estimated to be 4.3\% from the {\sc Treps} generator~\cite{TREPS}, while the one of 
$J=0$ is ignored with the decay to $\gpsp$ isotropic and no uncertainty in helicity. Belle measures the luminosity with 
1.4\% precision. The trigger efficiency for the events surviving the selection criteria exceeds 99.4\%, and so the 
uncertainty is ignored. The uncertainties of the $\jpsi$ and $\psp$ decay branching fractions taken from Ref.~\cite{PDG} 
contribute a systematic uncertainty of 1.3\%. The statistical error in the MC determination of the efficiency is less 
than 0.7\%.

The non-fit-related systematic uncertainties are listed in Table~\ref{sys_err}. Assuming all the sources are independent, 
we add them in quadrature to obtain a total systematic uncertainty of 6.2\% (4.4\%) of $J=2$ ($J=0$) in determining 
$\BR\cdot\Gamma_{\GG}$, in addition to the uncertainties from the fits. 

\begin{table}[tbp]
\centering
\begin{tabular}{| c | c  c|}
\hline 
Source & \multicolumn{2}{c|}{Relative error (\%)} \\\hline 
--- & $J=0$ & $J=2$ \\\hline
Particle identification     & \multicolumn{2}{c|}{2.8} \\ 
Tracking efficiency         & \multicolumn{2}{c|}{1.4} \\ 
Photon reconstruction       & \multicolumn{2}{c|}{2.0} \\ 
$\psp$ mass window          & \multicolumn{2}{c|}{0.6} \\ 
$P^*_t(\psp)$ and $P_t^*(\gpsp)$ & \multicolumn{2}{c|}{1.0} \\
$\MMS(\gpsp)$               & \multicolumn{2}{c|}{0.5} \\ 
Luminosity                  & \multicolumn{2}{c|}{1.4} \\ 
Generator                   & --- & 4.3 \\ 
Branching fractions         & \multicolumn{2}{c|}{1.3} \\
Statistics of MC samples    & \multicolumn{2}{c|}{0.7} \\ \hline
Sum in quadrature           & 4.4 &6.2 \\ \hline 
\end{tabular} 
\caption{The summary of systematic uncertainties besides the fits in $\GG\to \gpsp$ measurement.} 
\label{sys_err} 
\end{table}

\section{Discussion on the two structures}

We find evidence for the structure $R_1$ near $3.92~\gevcs$, which may be $X(3915)$, $\chi_{c2}(3930)$, or an admixture 
of them. Assuming $R_1$ is $\chi_{c2}(3930)$ and taking into account $\Gamma_{\GG}\BR(\chi_{c2}(3930)\to\ddb) = 210\pm 
40~\ev$, the ratio $R = \BR(\chi_{c2}(3930)\to\gpsp)/\BR(\chi_{c2}(3930)\to\ddb) = 0.010\pm 0.003$ is obtained. A rough 
estimation shows the partial width $\Gamma(\chi_{c2}(3930)\to\gpsp) = (200\sim 300)~\kev$, which is close to the 
predicted value of $207~\kev$ from Godfrey-Isgur relativistic potential model~\cite{theo}. 

It is interesting to see that the mass of $R_2$ agrees with the HQSS-predicted mass ($\approx 4013~\mevcs$) of the 
$2^{++}$ partner of $X(3872)$~\cite{x2-guofk}. The mass difference between $R_2$ and $X(3872)$ is $142.6\pm 4.2~\mevcs$, 
while that between $D^{*0}(2007)$ and $D^0$ is $142.01~\mevcs$. Meanwhile, the width of $R_2$ from the fit coincides 
with the predicted width of $2-8~\mevcs$ for the $2^{++}$ partner of $X(3872)$~\cite{x2-width-guofk}. Thus, $R_2$ may 
provide important information for understanding the nature of the $X(3872)$. However, the global significance of $R_2$ 
is only $2.8\sigma$. A much larger data sample that will be collected by Belle II may resolve this in the near future.

\section{Summary}

The two-photon process $\GG\to \gpsp$ is studied in the $\GG$ mass range from the threshold to $4.2~\gevcs$ for the 
first time with the full Belle data sample, and two structures are seen in the invariant mass distribution of $\gpsp$. 
The first has a mass of $M_{R_1} = 3922.4\pm 6.5 \pm 2.0~\mevcs$ and a width of $\Gamma_{R_1} = 22\pm 17 \pm 4 ~\mev$ 
with a local statistical significance of $3.1\sigma$ when the systematic uncertainties are included. This is close to 
the mass of $X(3915)$ and $\chi_{c2}(3930)$. The second has a mass of $M_{R_2} = 4014.3\pm 4.0 \pm 1.5~\mevcs$ and a 
width of $\Gamma_{R_2} = 4\pm 11 \pm 6 ~\mev$, with a global statistical significance of $2.8\sigma$. The values of 
$\Gamma_{\GG}\BR(R\to \gpsp)$ are of the order of several $\ev$.

\acknowledgments

We thank the KEKB group for the excellent operation of the accelerator; the KEK cryogenics group for the efficient
operation of the solenoid; and the KEK computer group, and the Pacific Northwest National Laboratory (PNNL) 
Environmental Molecular Sciences Laboratory (EMSL) computing group for strong computing support; and the National
Institute of Informatics, and Science Information NETwork 5 (SINET5) for valuable network support.  We acknowledge 
support from the Ministry of Education, Culture, Sports, Science, and Technology (MEXT) of Japan, the Japan Society 
for the Promotion of Science (JSPS), and the Tau-Lepton Physics Research Center of Nagoya University; 
the Australian Research Council including grants
DP180102629, % Sevior
DP170102389, % Varvell
DP170102204, % Yabsley
DP150103061, % Urquijo
FT130100303; % Urquijo;
Austrian Federal Ministry of Education, Science and Research (FWF) and FWF Austrian Science Fund No.~P~31361-N36;
the National Natural Science Foundation of China under Contracts
No.~11435013,  %Zhen-An Liu
No.~11475187,  %Chang-Zheng Yuan
No.~11521505,  %Chang-Zheng Yuan
No.~11575017,  %Cheng-Ping Shen
No.~11675166,  %Wen-Biao Yan
No.~11705209;  %Yi-Ming Li
No.~12175041;  %Xiaolong Wang
Key Research Program of Frontier Sciences, Chinese Academy of Sciences (CAS), Grant No.~QYZDJ-SSW-SLH011; % Chang-Zheng
the  CAS Center for Excellence in Particle Physics (CCEPP); %Chang-Zheng Yuan,
the Shanghai Science and Technology Committee (STCSM) under Grant No.~19ZR1403000; %Xiaolong Wang
the Ministry of Education, Youth and Sports of the Czech Republic under Contract No.~LTT17020;
Horizon 2020 ERC Advanced Grant No.~884719 and ERC Starting Grant No.~947006 ``InterLeptons'' (European Union);
the Carl Zeiss Foundation, the Deutsche Forschungsgemeinschaft, the Excellence Cluster Universe, and the 
VolkswagenStiftung; the Department of Atomic Energy (Project Identification No. RTI 4002) and the Department of 
Science and Technology of India; the Istituto Nazionale di Fisica Nucleare of Italy; 
National Research Foundation (NRF) of Korea Grant Nos.~2016R1\-D1A1B\-01010135, 2016R1\-D1A1B\-02012900, 
2018R1\-A2B\-3003643, 2018R1\-A6A1A\-06024970, 2018R1\-D1A1B\-07047294, 2019K1\-A3A7A\-09033840,
2019R1\-I1A3A\-01058933;
Radiation Science Research Institute, Foreign Large-size Research Facility Application Supporting project, the Global 
Science Experimental Data Hub Center of the Korea Institute of Science and Technology Information and KREONET/GLORIAD;
the Polish Ministry of Science and Higher Education and the National Science Center;
the Ministry of Science and Higher Education of the Russian Federation, Agreement 14.W03.31.0026, % from 15.02.2018
and the HSE University Basic Research Program, Moscow; % from 15.04.2021
University of Tabuk research grants
S-1440-0321, S-0256-1438, and S-0280-1439 (Saudi Arabia);
the Slovenian Research Agency Grant Nos. J1-9124 and P1-0135;
Ikerbasque, Basque Foundation for Science, Spain;
the Swiss National Science Foundation; 
the Ministry of Education and the Ministry of Science and Technology of Taiwan;
and the United States Department of Energy and the National Science Foundation.

\end{document}